\documentclass[pre,nofootinbib,superscriptaddress,twocolumn,notitlepage]{revtex4-1}

\usepackage[]{amsmath}  			
\usepackage{amssymb}    			
\usepackage{array}      			
\usepackage{comment}
\usepackage{color}
\usepackage{xcolor}

\usepackage{graphicx}           	
\usepackage[update]{epstopdf}   	

\newcommand{\trpa}[1]{{\textcolor{black}{#1}}}
\newcommand{\trp}[1]{{{#1}}}

\makeatletter
\DeclareOldFontCommand{\rm}{\normalfont\rmfamily}{\mathrm}
\DeclareOldFontCommand{\sf}{\normalfont\sffamily}{\mathsf}
\DeclareOldFontCommand{\tt}{\normalfont\ttfamily}{\mathtt}
\DeclareOldFontCommand{\bf}{\normalfont\bfseries}{\mathbf}
\DeclareOldFontCommand{\it}{\normalfont\itshape}{\mathit}
\DeclareOldFontCommand{\sl}{\normalfont\slshape}{\@nomath\sl}
\DeclareOldFontCommand{\sc}{\normalfont\scshape}{\@nomath\sc}
\makeatother

\begin{document}

\def\ldotsplus{\mathinner{\ldotp\ldotp\ldotp\ldotp}}
\def\fourdots{\relax\ifmmode\ldotsplus\else$\m@th \ldotsplus\,$\fi}

\title{Tunable wrinkling of thin nematic liquid crystal elastomer sheets}

\author{Madison S. Krieger}
\email{mkrieger@fas.harvard.edu}
\affiliation{Program for Evolutionary Dynamics, Harvard University, Cambridge, MA 02138, USA.}

\author{Marcelo A. Dias}
\email{madias@eng.au.dk}
\affiliation{Department of Engineering, Aarhus University, Inge Lehmanns Gade 10, 8000 Aarhus C, Denmark.}
\affiliation{Aarhus University Centre for Integrated Materials Research---$\mathrm{iMAT}$, Ny Munkegade 120, 8000 Aarhus C, Denmark}

\date{\today}

\makeatother

\begin{abstract}
Instabilities in thin elastic sheets, such as wrinkles, are of broad interest both from a fundamental viewpoint and also because of their potential for engineering applications. Nematic liquid crystal elastomers offer a new form of control of these instabilities through direct coupling between microscopic degrees of freedom, resulting from orientational ordering of rod-like molecules, and macroscopic strain. By a standard method of dimensional reduction, we construct a plate theory for thin sheets of nematic elastomer. We then apply this theory to the study of the formation of wrinkles due to compression of a thin sheet of nematic liquid crystal elastomer atop an elastic or fluid substrate. We find the scaling of the wrinkle wavelength in terms of material parameters and the applied compression. The wavelength of the wrinkles is found to be non-monotonic in the compressive strain owing to the presence of the nematic. Finally, due to soft modes, the critical stress for the appearance of wrinkles can be much higher than in an isotropic elastomer and depends nontrivially on the manner in which the elastomer was prepared. 
\end{abstract}






\maketitle

\section{Introduction}

The wrinkling of a thin film is a common occurrence, but elucidation of its underlying physics has far-reaching implications~\cite{Li_etal2012}. These include suggesting better ways to fabricate functional surfaces that change shape in predictable ways when subject to appropriate stimuli~\cite{TokarevMilko2008}, finding new and more accurate methods for measuring material properties~\cite{Wang2009,Chung_etal2011}, determining growth and form in biological tissues~\cite{thompson1942,LiangMaha2001}, and bringing insights into the fundamental mechanisms of pattern formation~\cite{bg2005,Davidovitch2011,Vella2011,King2012,Paulsen2015}. Many recent studies have focused on how wrinkling arises in isotropic materials as they respond to external effects, for example, when tension is applied to a freely suspended thin sheet~\cite{CerdaRavi-ChandarMahadevan2002,CerdaMahadevan2003,GeminardBernalMelo2004}, from compressing a film atop a soft substrate~\cite{ChenHutchinson2004,HuangHongSuo05,AudolyBoudaoud2008} or a fluid interface~\cite{Huang_etal2010,Brau2013,Oshri2015}, or through swelling mechanisms~\cite{GuvendirenYangBurdick2009}. However, less attention has been given to wrinkling instabilities in anisotropic materials, such as Nematic Liquid Crystal Elastomers (NLCE)~\cite{kf1991,WarnerTerentjev}.

NLCE are promising materials for studying and controlling surface pattern formation~\cite{wb2015,mwww2016,gymss2017,ahyxky2018,wm2018,kmgww2018,Lavrentovich2018,btghwsbl2018} due to unique effects that arise from the coupling between strain and nematic order~\cite{kf1994,uht2006,rkm2018}. For example, NLCE can display elastic soft modes, in which they can undergo deformation with negligible energy cost~\cite{KundlerFinkelmann1995,wbsww2016}. These materials, therefore, represent an extension of traditional anisotropic plates~\cite{rr2004,Lekhnitsky1957}, where the angle of anisotropy is now free to vary while the plate remains fixed. It has been shown that, under tension, thin sheets of NLCE can also develop intricate microstructures which suppress pattern formation at larger scales~\cite{pb2017}. Because the nematic phase is easily affected by electromagnetic fields, strain-order coupling also allows for nuanced actuation of these materials via light~\cite{dgn2016,tptw2018,gskabrw2018,awhkbgbdw2018}, and one can also measure and record their deformation electronically~\cite{rkbaspw2018}. They are also an excellent candidate for artificial muscle tissue~\cite{mp2018}. 

\begin{figure}
\includegraphics[width=0.9\columnwidth]{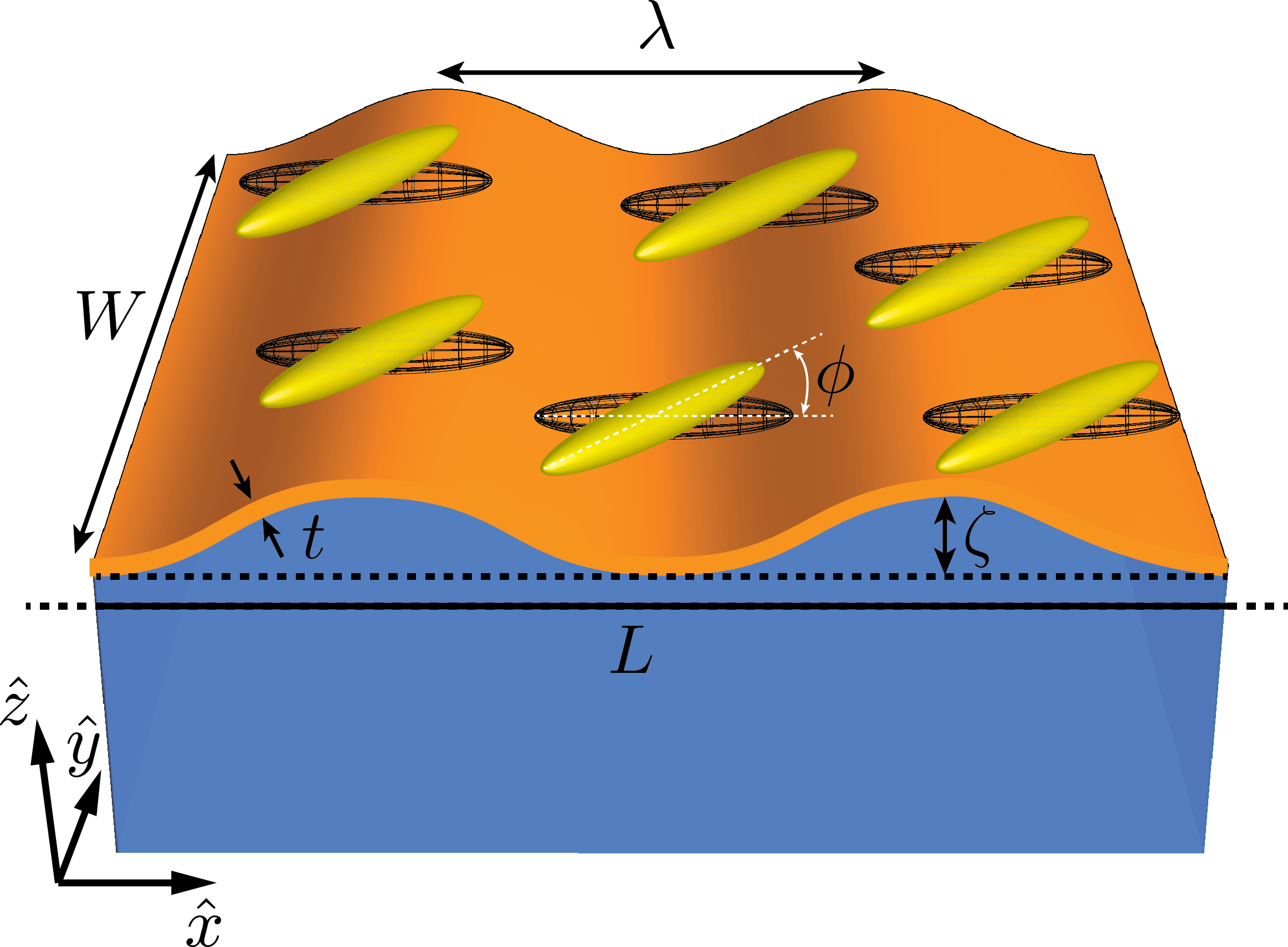}
\caption{\footnotesize A cartoon is used to represent the geometry and the parameters of the problem. A NLCE plate of thickness $t$, width $W$ and length $L$ is bonded to a substrate, which may be fluid or elastic. The bonded pair are compressed by a percentage $\gamma$ along the longitudinal direction, which we take to be $\hat{x}$. The nematic director is assumed to lie in the tangent plane of the thin film. The director is parameterized after dimensional reduction by the angle $\phi$ measured in the tangent plane, and the out-of-plane deformation is $\zeta(x,y)$. }
\label{setupf}
\end{figure}

Although the mechanics of NLCE has been considered by several authors~\cite{WarnerMahadevan2004,WarnerModesCorbett2010,AharoniSharonKupferman2014,CirakWarner2014,ad2017,ko2017,Nguyen2017}, a first principle derivation of a dimensionally reduced model from an effective theory for these materials in the limit of thin plates is still lacking. In this article we derive a F\"{o}ppl-von K\'{a}rm\'{a}n-like plate theory for NLCE and use this theory to study the wrinkling of such thin materials under compressive loads. The specific problem to be addressed is when a thin sheet of NLCE is placed on a soft isotropic elastic foundation, or a fluid sub-phase. Our problem is thus an inversion of already existent experiments and theory on wrinkling of an isotropic plate atop a thick nematic elastomer foundation~\cite{palffymuhorayNLCE,Soni2016}. As depicted in Fig.~\ref{setupf}, the wrinkling of the plate atop the foundation is induced by compressing both the foundation and the plate as a unit. In the absence of a foundation, the plate will choose an out-of-plane deformation which is a single arch; if the plate is bonded to the foundation, however, this deformation has a high energetic cost, due either to deformation of an elastic foundation or the gravitational energy stored in a fluid substrate. 

In Sections \ref{3densec} and \ref{dimredsec}, we outline the dimensional reduction. We begin with a three-dimensional phenomenological energy and extend the kinematic and dynamic Kirchhoff-Love approximations with compatible approximations for the behavior of the nematic director. These approximations allow us to integrate across the thickness of the plate, thus reducing the three-dimensional energy to a two-dimensional energy comprised of two terms---a stretching-like term proportional to the thickness $t$, and a bending-like term proportional to $t^3$. These terms couple the geometry of the macroscopic sheet to the director orientation. We explore the ramifications of our model by considering a common wrinkling \emph{ansatz}, in which a base state of stress for a compressed but planar plate is energetically coupled to the strain resulting from out-of-plane displacement. The base state of stress is derived in Section \ref{inplanesec}. The approximations for the bending and stretching energies due to the \emph{ansatz} are discussed in Section \ref{outplanesec}, where the scaling of the wrinkle wavelength is also considered. The scaling has the same functional form as for an isotropic plate, but includes information from the nematogen-elastomer coupling strengths. We summarize our results in Section \ref{sumsec}. 

\section{Three-dimensional energy}
\label{3densec}

We assume that the elastomer is prepared in such a way that the material was deep in the nematic phase at the time of cross-linking, and denote by $\bar{\mathbf{n}}$ the unit vector director field in this initial configuration, also known as the ``reference" configuration. This assumption is important in NLCE because there is an extra term in the energy for the elastomer depending on whether the material was prepared and cross-linked with the liquid crystal in the isotropic phase~\cite{Lubensky_etal2002,WarnerTerentjev}. When the NLCE is subject to external forces or torques, there exist\trp{s} an energetic cost to deform the elastomer and rotate the director field to a new configuration $\mathbf{n}$. To describe this energy, we review standard notions from continuum mechanics.

\trp{The reference configuration of the elastomer is a thin slab. We define an orthonormal basis $\{\mathbf{e}_i\}$ with $\mathbf{e}_{1}$ and $\mathbf{e}_2$ parallel to the plane of the slab, and $\mathbf{e}_3$ perpendicular to the slab. Cartesian coordinates $x_1$, $x_2$, $x_3$ determine the positions of points relative to this frame. The $x_i$ will also serve as Lagrangian coordinates for the deformed configuration, where the position of a material point $p$, originally at $\mathbf{x}$, is given by the deformation map $\mathbf{Y}(\mathbf{x})$. The deformation map defines another natural basis $\{\mathbf{g}_i\}$, where $\mathbf{g}_i=\partial_i\mathbf{Y}\equiv\partial\mathbf{Y}/\partial x_i.$ Due to the deformation, the distance between material points changes; the distance between material points is encoded in the metric tensor \trpa{$\boldmath{\mathsf{g}}=g_{ij}\mathbf{e}^i\otimes\mathbf{e}^j$, where} $g_{ij}=\mathbf{g}_i\cdot\mathbf{g}_j$. Denoting the inverse of the metric tensor by $g^{ij}$, we can define the dual basis $\{\mathbf{g}^i\}$ via $\mathbf{g}^i=g^{ij}\mathbf{g}_j$ so that $\mathbf{g}^i\cdot\mathbf{g}_j=\delta^i_j$. (Note that we use the Einstein convention of summing repeated indices.)  For a deformed configuration, $\mathbf{g}^i\neq\mathbf{g}_i$, but $\mathbf{e}^i=\mathbf{e}_i$ for the Cartesian basis. The deformation tensor  $\mathsf{F}$ maps the Cartesian basis vectors $\mathbf{e}_i$ onto the deformed basis vectors $\mathbf{g}_i$, i.e. $\mathsf{F}:\mathbf{e}_i\mapsto\mathbf{g}_i$. Thus, $\mathsf{F}=\mathbf{g}_i\otimes\mathbf{e}^i$, where $\otimes$ denotes the outer product. Note that $\mathsf{F}=\nabla\mathbf{Y}=\partial_jY_i\mathbf{e}^i\otimes\mathbf{e}^j$, \trpa{or $F_{ij}=\partial_j Y_i$}. We can define the Lagrangian or Green strain tensor  $\boldsymbol\varepsilon$ in terms of reference-space quantities \trpa{$\mathrm{d}Y_i\mathrm{d}Y_i-\mathrm{d}x_i\mathrm{d}x_j=2\varepsilon_{ij}\mathrm{d}x_i\mathrm{d}x_j$}, and the Eulerian or  Almansi strain tensor $\boldsymbol\varepsilon^*$ in terms of deformed- or ``target"-space quantities, \trpa{$\mathrm{d}Y_i\mathrm{d}Y_i-\mathrm{d}x_i\mathrm{d}x_j=2\varepsilon^*_{ij}\mathrm{d}Y_i\mathrm{d}Y_j$:}}
\begin{subequations}
\label{strainAG}
	\begin{align}
		\boldsymbol\varepsilon&= \frac{1}{2}\left[\mathsf{F}^{\mathrm{T}}\mathsf{F}-\mathsf{1}\right]=\varepsilon_{ij}\mathbf{e}^i\otimes\mathbf{e}^j
		\label{strainG} \\
		\boldsymbol\varepsilon^*&= \frac{1}{2}\left[\mathsf{1}^*-\mathsf{F}^{\mathrm{-T}}\mathsf{F}^{-1}\right]=\varepsilon^*_{ij}\mathbf{g}^i\otimes\mathbf{g}^j,
		\label{strainA}
	\end{align}
\end{subequations}
where $\mathsf{1}=\mathbf{e}_i\otimes\mathbf{e}^i$ \trp{is the identity on the reference space} and $\mathsf{1}^*=\mathbf{g}_i\otimes\mathbf{g}^i$ \trp{is the identity on the target space}.

The resistance of the cross-linked network to deformation is described by an energy density which is quadratic in the strain. The coupling between the orientational order and the strain of the polymer network is described by de Gennes' phenomenological energy~\cite{DeGennes1975,Mbanga2010,Sawa2011}:
\begin{eqnarray}
\label{eq:EnergyTotal21}
		{E}&=&\frac{1}{2}\int\mathrm{d}V\left[\lambda\left(\mathrm{Tr}\boldsymbol\varepsilon\right)^2+2\mu\mathrm{Tr}\left(\boldsymbol\varepsilon^2\right) \right.\nonumber\\
&&\left.-2{\alpha}\left(\mathbf{n}\cdot\boldsymbol\varepsilon^*\cdot\mathbf{n}-\bar{\mathbf{n}}\cdot\boldsymbol\varepsilon\cdot\bar{\mathbf{n}}\right)+hf(\bar{\mathbf{n}},\mathbf{n})\right],
\end{eqnarray}
where $\lambda$ and $\mu$ are the Lam\'{e} coefficients, and $\alpha$ and $h$ are coupling constants. The term proportional to $\alpha$ is the lowest-order term involving the nematic directors and strains. Since the reference director $\bar{\mathbf{n}}$ is a vector under the rotation group in the reference space, it must couple with the Green strain tensor in order to make the energy density a scalar~\cite{Lubensky_etal2002}, which gives the term $\bar{\mathbf{n}}\cdot\boldsymbol\varepsilon\cdot\bar{\mathbf{n}}$. Similarly, the deformed director $\mathbf{n}$ must couple with the Almansi strain tensor, thus $\mathbf{n}\cdot\boldsymbol\varepsilon^*\cdot\mathbf{n}$. The thermomechanical history of the sample is captured by the term proportional to $h$~\cite{Kenji_history}. For materials prepared deep in the nematic phase, the cross-links themselves resist rotating away from the reference state and, therefore, they ``remember'' the orientation of the nematic at the time of cross-linking~\cite{WarnerTerentjev}. A low value of $h$ corresponds to a soft nematic elastomer, which has a low cross-link density, whereas a high value of $h$ corresponds to a nematic glass with high cross-link density. The function $f(\bar{\mathbf{n}},\mathbf{n})$ captures the energetic cost of rotating the director relative to the polymer matrix background. The symmetries also restrict the possible forms of the function $f(\bar{\mathbf{n}},\mathbf{n})$, since it must behave as a scalar under rotations in both the reference and deformed spaces. Since $\bar{\mathbf{n}}=\bar{n}_i\mathbf{e}^i$ and $\mathbf{n}=n_i\mathbf{g}^i$, we choose to construct $f(\bar{\mathbf{n}},\mathbf{n})$ by using the deformation map in such a way that $\mathbf{n}$ is mapped back to the reference space. Therefore, we define $f(\bar{\mathbf{n}},\mathbf{n})\equiv \mathrm{Tr}\Delta\mathsf{Q}^2$, where $\Delta\mathsf{Q}\equiv\left(\mathsf{F}^{-1}\mathbf{n}\right)\otimes\left(\mathsf{F}^{-1}\mathbf{n}\right)-\bar{\mathbf{n}}\otimes\bar{\mathbf{n}}$.

A derivation of the coupling between strain and orientation order starting with the neoclassical NLCE energy~\cite{WarnerTerentjev} reveals that $0<\alpha<3\mu/2$~\cite{UchidaOnuki1999}. The only restriction on $h$ is that it must be positive.

\section{Dimensional Reduction}
\label{dimredsec}

The dimensional reduction procedure considered here follows standard methods available for isotropic thin plates or shells~\cite{Koiter1970,Ciarlet2005}. We assume that the volume element may be decomposed into $\mathrm{d}V=\mathrm{d}\mathcal{A}^{\mbox{\tiny(0)}}\,\mathrm{d}{x_3}$, where $\mathrm{d}\mathcal{A}^{\mbox{\tiny(0)}}=\mathrm{d}{x_1}\mathrm{d}{x_2}$ is the area element of the middle reference-plane and $\mathrm{d}{x_3}$ is the integration measure through the thickness, ${x_3}\in[-t/2,t/2]$. Furthermore, simplifications are made if we assume two \emph{a priori} conditions~\cite{John1965,Koiter1970,Ciarlet2005}: (i) a mechanical condition, which states that the state of stress inside the body as purely parallel to the mid-surface; and (ii) a kinematic condition, known as the Kirchhoff-Love condition, which states that points located along the normal to the middle plane remain along the normal after the deformation to an arbitrary surface, while their distance with respect to this mid-surface does not change. In order to complement (i) and (ii) in light of the problem at hand, we impose a third condition, which is a kinematic constraint on the distribution of nematic directors: (iii) on every surface of constant $x_3$, we shall assume that both the reference and deformed director field remain tangent to the mid-surface of the film, i.e. $\bar{n}_3=0$ and $n_3=0$. Note that this last equality applies to the director in the deformed space.

The embedding of the plate in the deformed configuration is explicitly given in normal coordinates by assuming that a point $p$ (such that $\mathbf{Y}:p=(x_1,x_2,x_3)\mapsto\mathbb{R}^3$) in the body is written in terms of a point in the mid-surface $p^{\mbox{\tiny(0)}}$  (such that $\mathbf{S}:p^{\mbox{\tiny(0)}}=(x_1,x_2)\mapsto\mathbb{R}^3$) through the relationship $p=p^{\mbox{\tiny(0)}}+x_3\mathbf{e}_3$, where $\mathbf{Y}(p^{\mbox{\tiny(0)}})=\mathbf{S}(p^{(0)})$. It is, therefore, a critical step in the dimensional reduction to express the embedding evaluated at $p$ as a Taylor series about points in the mid-surface:
\begin{equation}
	\mathbf{Y}(p)=\mathbf{S}(p^{\mbox{\tiny(0)}})+x_3\mathsf{F}^{\mbox{\tiny(0)}}\mathbf{e}_3+\cdots,
\end{equation} 
were we have used the definition $\mathsf{F}^{\mbox{\tiny(0)}}\equiv\left.\nabla\mathbf{Y}\right|_{p^{\mbox{\tiny(0)}}}$. We note that $\mathsf{F}^{\mbox{\tiny(0)}}$ is the deformation gradient of the mid-surface and, given the corresponding basis set on the mid-surface $\mathbf{a}_i=\left\{\mathbf{a}_\alpha,\mathbf{a}_3\right\}$, we may write  $\mathsf{F}^{\mbox{\tiny(0)}}=\mathbf{a}_\alpha\otimes\mathbf{e}^\alpha+\mathbf{a}_3\otimes\mathbf{e}^3$, where $\{\alpha,\beta,\cdots\}$ take values in the set $\{1,2\}$. An additional clarification is that the basis set is consistently defined using the mid-surface embedding, \emph{i.e.} $\mathbf{a}_\alpha\equiv\partial_\alpha\mathbf{S}$ and $\mathbf{a}^3=\mathbf{a}_3=\mathbf{a}_1\times\mathbf{a}_2/|\mathbf{a}_1\times\mathbf{a}_2|$. \trp{We can relate basis vectors $\mathbf{g}_i\equiv\partial_i\mathbf{Y}$ on surfaces of constant $x_3$ to the corresponding basis vectors $\mathbf{a}_i$ on the mid surface, and vice-versa, via a translation tensor $\mathsf{T}:\mathbf{a}_i\mapsto\mathbf{g}_i$ which is given by $\mathsf{T}=\mathbf{g}_i\otimes\mathbf{a}^i$ (and $\mathsf{T}^{-1}=\mathbf{a}_i\otimes\mathbf{g}^i$)}~\cite{Pietraszkiewicz1980}. Through the definition of the mid-surface metric tensor, $\mathsf{a}=a_{\alpha\beta}\mathbf{a}^\alpha\otimes\mathbf{a}^\beta=\mathbf{a}_\alpha\otimes\mathbf{a}^\alpha$, and curvature tensor, $\mathsf{b}=b_{\alpha\beta}\mathbf{a}^\alpha\otimes\mathbf{a}^\beta=-\partial_\alpha\mathbf{a}^3\otimes\mathbf{a}^\alpha$, we may write $\mathsf{T}=\mathsf{1}-x_3\mathsf{b}$ ($\mathsf{T}^{-1}=\mathsf{1}+x_3\mathsf{b}+\cdots$). From the above definition, we may also apply the translation tensor to the deformation gradient, which results in the relation $\mathsf{F}=\mathsf{T}\mathsf{F}^{\mbox{\tiny(0)}}$. The above definitions allow us to write both strain tensors, in terms of their components, as follows:
\begin{equation}
\boldsymbol\varepsilon=\boldsymbol\varepsilon^{\mbox{\tiny(0)}}-x_3\mathsf{F}^{\mbox{\tiny(0)}\mathrm{T}}\mathsf{b}\mathsf{F}^{\mbox{\tiny(0)}}+\frac{1}{2}x_3^2\mathsf{F}^{\mbox{\tiny(0)}\mathrm{T}}\mathsf{c}\mathsf{F}^{\mbox{\tiny(0)}},
		\label{strainG2}
\end{equation}
where
\begin{subequations}
\label{FF}
	\begin{align}
		\boldsymbol\varepsilon^{\mbox{\tiny(0)}}=\frac{1}{2}\left[\mathsf{F}^{\mbox{\tiny(0)}\mathrm{T}}\mathsf{F}^{\mbox{\tiny(0)}}-\mathsf{1}\right]=\frac{1}{2}\left(a_{\alpha\beta}-\delta_{\alpha\beta}\right)\mathbf{e}^\alpha\otimes\mathbf{e}^\beta,
		\label{IFF} \\
		\mathsf{F}^{\mbox{\tiny(0)}\mathrm{T}}\mathsf{b}\mathsf{F}^{\mbox{\tiny(0)}}=b_{\alpha\beta}\mathbf{e}^\alpha\otimes\mathbf{e}^\beta,
		\label{IIFF} \\
		\mathsf{F}^{\mbox{\tiny(0)}\mathrm{T}}\mathsf{c}\mathsf{F}^{\mbox{\tiny(0)}}=c_{\alpha\beta}\mathbf{e}^\alpha\otimes\mathbf{e}^\beta,
		\label{IIIFF}
	\end{align}
\end{subequations}
with $c_{\alpha\beta}=\partial_\alpha\mathbf{a}^3\cdot\partial_\beta\mathbf{a}^3$ are the components of the third fundamental form. Also,
\begin{eqnarray}
	\boldsymbol\varepsilon^*&=\mathsf{T}^{-1}\left(\boldsymbol\varepsilon^{\mbox{\tiny(0)}*}-x_3\mathsf{b}+\frac{1}{2}x_3^2\mathsf{c}\right)\mathsf{T}^{-1}\nonumber\\
	&=\boldsymbol\varepsilon^{\mbox{\tiny(0)}*}-x_3\mathsf{b}+\frac{1}{2}x_3^2\mathsf{c}+\cdots,
		\label{strainA2}
\end{eqnarray}
where higher-order terms in $x_3$ have been neglected, and 
\begin{subequations}
\label{FFA}
	\begin{align}
		\!\!\!\!\boldsymbol\varepsilon^{\mbox{\tiny(0)}*}\!=\!\frac{1}{2}\left[\mathsf{1}^*\!\!-\!\mathsf{F}^{\mbox{\tiny(0)}\mathrm{-T}}\mathsf{F}^{\mbox{\tiny(0)}-1}\right]\!\!=\!\frac{1}{2}\left(a_{\alpha\beta}-\delta_{\alpha\beta}\right)\mathbf{a}^\alpha\otimes\mathbf{a}^\beta
		\label{IFFA} \\
		\mathsf{b}=b_{\alpha\beta}\mathbf{a}^\alpha\otimes\mathbf{a}^\beta
		\label{IIFFA} \\
		\mathsf{c}=c_{\alpha\beta}\mathbf{a}^\alpha\otimes\mathbf{a}^\beta.
		\label{IIIFFA}
	\end{align}
\end{subequations}
From the above expressions, it is clear that $\boldsymbol\varepsilon^*=\mathsf{F}^{\mbox{\tiny(0)}\mathrm{T}}\boldsymbol\varepsilon\mathsf{F}^{\mbox{\tiny(0)}}$ is satisfied. Despite differences in the Eulerian and Lagrangian formulations~\cite{Pietraszkiewicz1980}, Eqs.~\eqref{strainG2} and \eqref{strainA2} show that these two approaches yield identical tensorial components, in terms of the components of the fundamental forms in the deformed mid-surface~\cite{doCarmo1976}, expressed with different bases. In practical terms, the mixed contribution $\mathbf{n}\cdot\boldsymbol\varepsilon^*\cdot\mathbf{n}-\bar{\mathbf{n}}\cdot\boldsymbol\varepsilon\cdot\bar{\mathbf{n}}$ results in contractions with the same components for the strain measure. Therefore, if we define $\varepsilon_{\alpha\beta}\equiv\left(a_{\alpha\beta}-\delta_{\alpha\beta}\right)/2-x_3b_{\alpha\beta}+x_3^2c_{\alpha\beta}/2$, the coupling terms become $\left(n^\alpha n^\beta-\bar{n}^\alpha\bar{n}^\beta\right)\varepsilon_{\alpha\beta}$.

Note that the curvature with a raised index is defined with the inverse of the deformed metric, $b^\beta_\alpha\equiv\left(\mathsf{a}^{-1}\right)^{\beta\delta}b_{\delta\alpha}$, which satisfies the property $\left(\mathsf{a}^{-1}\right)^{\alpha\gamma}a_{\gamma\beta}=\delta^\alpha_\beta$. However, by definition, we may write $a_{\alpha\beta}=\delta_{\alpha\beta}+2\varepsilon_{\alpha\beta}$, which through expanding in orders of the stain results in $\left(\mathsf{a}^{-1}\right)^{\alpha\beta}=\delta_{\alpha\beta}+\mathcal{O}\left(|\varepsilon|\right)$. Since the energy is already second-order in strain, we here assume the approximation that the Cartesian metric components $\delta_{\alpha\beta}$ raises and lowers the indices for the contractions in the energy, thus allowing us to use either covariant or contravariant notation, knowing that their difference only yields higher order contributions.

From the definition of the components of the stress tensor as 
\begin{equation}
	\sigma_{ij}=\delta {E}/\delta\varepsilon_{ij}=\lambda\delta_{ij}\varepsilon_{kk}+2\mu\varepsilon_{ij}-\alpha\Delta Q_{ij},
\end{equation} 
the condition (i) reads $\sigma_{33}=\sigma_{23}=\sigma_{13}=0$~\cite{landau_lifshitz_elas}. This step automatically implies that the third component of the three-dimensional strain tensor follows the constraint given by $\varepsilon_{33}=-\varepsilon_{\alpha\alpha}\lambda/(\lambda+2\mu)$, which allows us to rewrite the energy given in Eq.~\eqref{eq:EnergyTotal21} entirely in terms of components on the surfaces of constant $x_3$.

The final assumption (iii) tells us that the director field only has components parallel to the surfaces of constant ${x_3}$. We can therefore uniquely express the field $\mathbf{n}$ in terms of its projection onto the mid-surface, here defined by $n^{\mbox{\tiny(0)}}_\alpha$, via the relationship
\begin{equation}
\label{eq:Map-director}
\mathbf{n}=n_\alpha\mathbf{g}^\alpha=\left[n_\alpha\left(\mathsf{T}^{-1}\right)^\alpha_\beta\right]\mathbf{a}^\beta\equiv n^{\mbox{\tiny(0)}}_\beta\mathbf{a}^\beta,
\end{equation}
which gives $n_\alpha= T_\alpha^\beta n^{\mbox{\tiny(0)}}_\beta$.

The dependence on $x_3$ coming from the nematic contribution to the energy is explicitly written though the following expression,
\begin{eqnarray}
	\label{eq:Order}
		n_\alpha n_\beta &=&n^{\mbox{\tiny(0)}}_\alpha n^{\mbox{\tiny(0)}}_\beta-{x_3}B_{\alpha\beta}+\frac{1}{2}x_3^2C_{\alpha\beta},
\end{eqnarray}
where Eq.~\eqref{eq:Map-director} has been used, $B_{\alpha\beta}\equiv n^{\mbox{\tiny(0)}}_\alpha n^{\mbox{\tiny(0)}}_\delta b_{\delta\beta}+ b_{\alpha\delta} n^{\mbox{\tiny(0)}}_\delta n^{\mbox{\tiny(0)}}_\beta$, and $C_{\alpha\beta}\equiv 2 b_{\alpha\delta} n^{\mbox{\tiny(0)}}_\delta n^{\mbox{\tiny(0)}}_\gamma b_{\gamma\beta}$ (the approximation $b^\beta_\alpha\approx b_{\alpha\beta}$ has been employed in these definitions). Substituting these expansions into Eq.~\eqref{eq:EnergyTotal21} yields an energy density with an explicit functional dependence on ${x_3}$. Therefore, we may eliminate this degree of freedom by integrating along the body's thickness. The non-zero contributions to the two-dimensional energy come only from even powers of ${x_3}$, because of the choice of symmetric limits of integration, $[-t/2,t/2]$.
In this derivation, the terms proportional to $t^2\varepsilon^{\mbox{\tiny(0)}}_{\alpha\beta}c_{\alpha\beta}$ and $t^2\varepsilon^{\mbox{\tiny(0)}}_{\alpha\beta}C_{\alpha\beta}$ shall be neglected, as they are $\mathcal{O}\left(t^2\zeta^4/L^6\right)$ or smaller, where $\zeta$ is the scale of the typical out-of-plane deflection of material points in the $x_3$ direction and $L$ is a lateral length-scale of the plate. We write the two-dimensional energy in the usual thin-plate sense, ${E}=(1/2)\int\mathrm{d}\mathcal{A}^{\mbox{\tiny(0)}}\left[t\,\mathcal{E}_\mathrm{s} +(t^3/12)\mathcal{E}_\mathrm{b} \right]$, where the modified energy densities are written as follows:
\begin{subequations}
	\begin{align}
		\mathcal{E}_\mathrm{s}&=2 \mu\left(\varepsilon^{\mbox{\tiny(0)}2}_{\alpha\alpha}+\varepsilon^{\mbox{\tiny(0)}2}_{\alpha\beta}\right)-2\alpha\varepsilon^{\mbox{\tiny(0)}}_{\alpha\beta}\Delta{Q}^{\mbox{\tiny(0)}}_{\alpha\beta}+h\Delta{Q}^{\mbox{\tiny(0)}2}_{\alpha\beta}\label{stretchingenergy} \\
		\mathcal{E}_\mathrm{b}&=2 \mu\left(b_{\alpha\alpha}^2+b_{\alpha\beta}^2\right)-2 \alpha \left(b_{\alpha\beta}B_{\alpha\beta}+\frac{1}{2}c_{\alpha\beta}\Delta{Q}^{\mbox{\tiny(0)}}_{\alpha\beta}\right)\nonumber\\
&\quad+h\left(B_{\alpha\beta}^2+C_{\alpha\beta}\Delta{Q}^{\mbox{\tiny(0)}}_{\alpha\beta}\right), \label{bendingenergy}
	\end{align}
\end{subequations}
where $\Delta{Q}^{\mbox{\tiny(0)}}_{\alpha\beta}\equiv n^{\mbox{\tiny(0)}}_\alpha n^{\mbox{\tiny(0)}}_\beta-\bar{n}^{\mbox{\tiny(0)}}_\alpha \bar{n}^{\mbox{\tiny(0)}}_\beta$ and $\varepsilon^{\mbox{\tiny(0)}}_{\alpha\beta}\equiv\left(a_{\alpha\beta}-\delta_{\alpha\beta}\right)/2$. This energy is expressed entirely in terms of quantities defined on the mid-surface. Since the bulk modulus is typically much larger than the shear modulus in rubbery materials, we have assumed that the elastomer is nearly incompressible, \emph{i.e.} $\lambda/(\lambda+2\mu)\rightarrow1$ (or $\varepsilon_{33}=-\varepsilon_{\alpha\alpha}$). The first terms in Eqs.~\eqref{stretchingenergy}-\eqref{bendingenergy} are the usual stretching and bending, respectively, arising in isotropic plate theory. The last two terms in Eq.~\eqref{stretchingenergy} are readily seen to be inherited from the three-dimensional form of the bulk energy~(\ref{eq:EnergyTotal21}), while the last two in Eq.~\eqref{bendingenergy} couple bending to the orientational order.

It is noteworthy that in the limit $h \rightarrow \infty$, a limit associated with nematic glasses~\cite{bigginswarner2008,CirakWarner2014}, our energy recovers the strong coupling between nematic defects and elastomer curvature~\cite{wmc2010,modes_warner2011}. In particular, for the energy to be bounded in this limit, we must have $B^2_{\alpha\beta}+C_{\alpha \beta} \Delta Q^{(0)}_{\alpha \beta}=\mathcal{O}(1/h)$, which implies that products of the director field and second fundamental form components are small at every point. Because $C_{\alpha\beta}\equiv 2 b_{\alpha\delta} n^{\mbox{\tiny(0)}}_\delta n^{\mbox{\tiny(0)}}_\gamma b_{\gamma\beta}$, in a defect-free texture where the nematic director $n^{\mbox{\tiny(0)}}$ never vanishes, the only way for the energy to be bounded is to have the second fundamental form $b_{\alpha\delta}$ vanish. This implies that there is no curvature of the sheet for large $h$, which has large implications for wrinkling (since wrinkles induce curvature in the sheet), which we discuss in Section \ref{outplanesec}. 

Next, we use our derived plate theory to address the particular example of a thin NLCE plate bonded to an elastic or fluid substrate and placed under compression. Following previous works, we first derive the in-plane stress in the absence of buckling. For an isotropic elastomer, this step is crucial to understand how the stress and therefore the stretching energy scales in a near threshold regime. For NLCE, it is in addition necessary to determine the director orientation, which affects both the in-plane stress and the effective bending modulus, and therefore ``tunes'' the wavelength and critical compression for wrinkles. Having derived these quantities, we then consider a standard \emph{ansatz} for the buckled shape, and solve for the amplitude, wavelength, and critical stress for buckling in both the cases of a fluid and elastic substrate. 

\section{In-plane stress and director orientation}
\label{inplanesec}

First, we describe the state of stress in a flat configuration. We now adopt a global frame $(\hat{x},\hat{y},\hat{z})$ with $z$ being the thin direction, see Fig.~\ref{setupf}. The thin elastomer is of infinite width $0 < y < \infty$ and finite length $0 < x < L$, with initial director $\bar{\mathbf{n}}^{(0)} = \hat{\mathbf{x}}$ (we perform the analogous finite-width calculation in the Appendix). The plate is subjected to a compressive strain $\varepsilon^{\mbox{\tiny(0)}}_{xx}=-\gamma$. We will assume the deformations are small and that the out-of-plane deflection $\zeta$ is identically zero. Therefore, we write $\varepsilon^{\mbox{\tiny(0)}}_{\alpha\beta}=(\partial_\alpha u_\beta+\partial_\beta u_\alpha)/2$. We will often plot quantities against strains up to $25\%$, but the theory is expected to be numerically accurate only for small strains. To determine the base state of stress, we must solve the first F\"{o}ppl-von K\'{a}rm\'{a}n equation $\nabla \cdot \boldsymbol\sigma^{\mbox{\tiny(0)}} = 0$, where the stress is given by $\sigma^{\mbox{\tiny(0)}}_{\alpha\beta}=\partial \mathcal{E}_\mathrm{s}/\partial\varepsilon^{\mbox{\tiny(0)}}_{\alpha\beta}$ using the stretching energy density in Eq.~\eqref{stretchingenergy}. 

\begin{figure}[t]
\includegraphics[width=1\columnwidth]{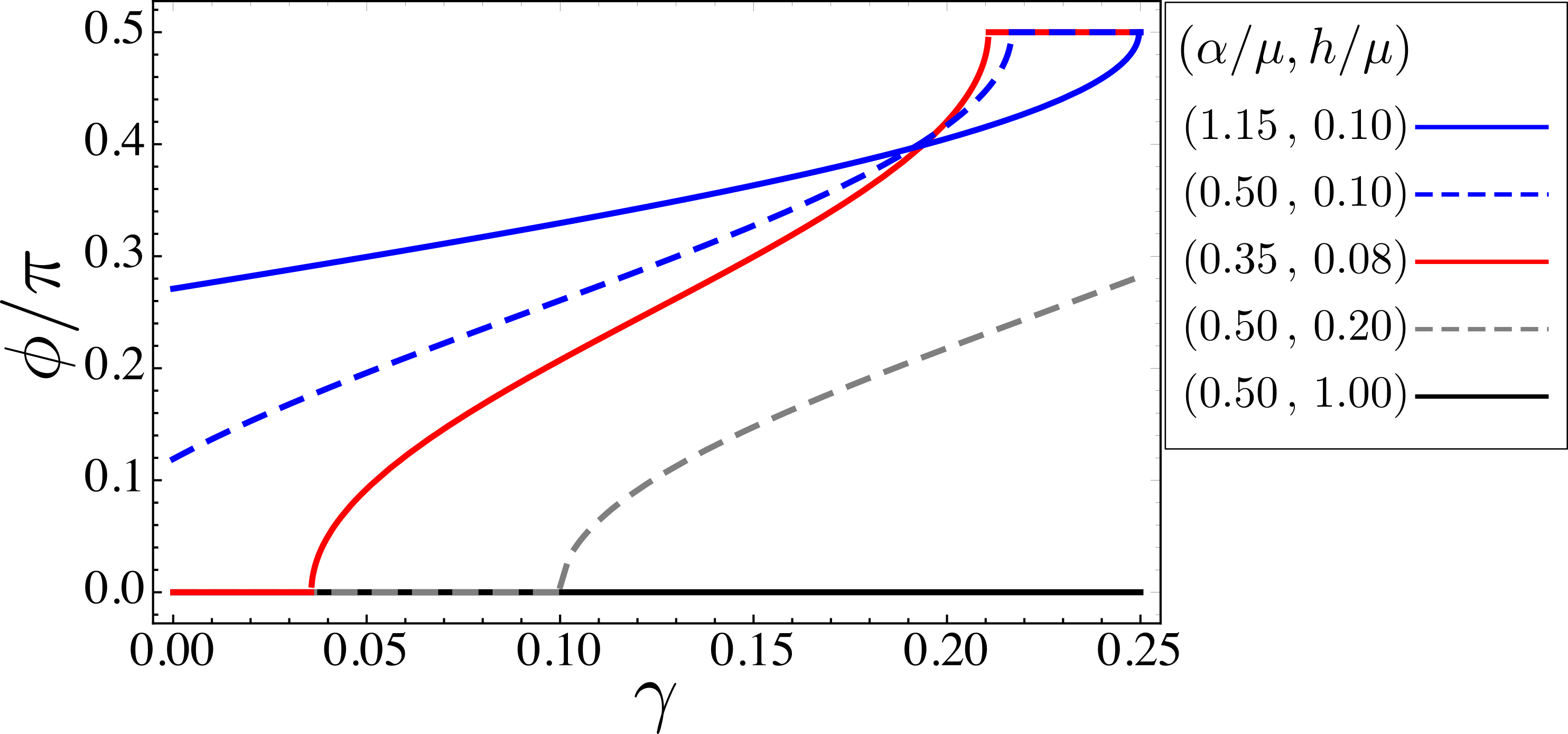}
\caption{\footnotesize Behavior of the director angle vs compressive strain $\gamma$ for different crosslinking strengths $h/\mu$ and strain-director couplings $\alpha/\mu$. For high $h/\mu$ and low $\alpha/\mu$ (black), the director will not rotate until a very high strain is reached. This critical strain becomes smaller as $h/\mu$ becomes smaller (gray-dashed). For $\alpha>h$, rotating the director becomes easier, and the director can even rotate a total of $\pi/2$ within a reasonable range of compressive strain (red). For increasing values of $\alpha/\mu$ (blue), the director rotates very quickly with any applied compressive strain $\gamma$, then more slowly until rotation is completed. For small $h/\mu$, there can be re-entrance as a function of $\alpha/\mu$ (red, blue-dashed, blue-solid).}
\label{tomographs}
\end{figure}

If $\phi$ is the angle between $\mathbf{n}^{\mbox{\tiny(0)}}$ and $\bar{\mathbf{n}}^{\mbox{\tiny(0)}}$, then these solutions are given by
\begin{subequations}
	\begin{align}
\sigma^{\mbox{\tiny(0)}}_{xx}&=2\mu\left(2\varepsilon^{\mbox{\tiny(0)}}_{xx}+\varepsilon^{\mbox{\tiny(0)}}_{yy}\right)+\alpha\sin^2\phi,\label{e1}\\
\sigma^{\mbox{\tiny(0)}}_{yy}&=2\mu\left(\varepsilon^{\mbox{\tiny(0)}}_{xx}+2\varepsilon^{\mbox{\tiny(0)}}_{yy}\right)-\alpha \sin^2\phi, \label{e2} \\
\sigma^{\mbox{\tiny(0)}}_{xy}&=2\mu\varepsilon^{\mbox{\tiny(0)}}_{xy}-\alpha\sin\phi\cos\phi. \label{e3}
	\end{align}
\end{subequations}
Note that due to the use of the rotation from the deformed space when defining $\mathbf{n}^{\mbox{\tiny(0)}}$, the angle $\phi$ may be different from what is measured in the laboratory frame. To convert to the angle measured in the lab frame, it is necessary to apply a local rotation obtained by taking the polar decomposition of the deformation tensor, which includes local metric information \cite{mbanga2012hybrid}. However, the observables of interest in this work are macroscopic geometric quantities, such as the wavelength of wrinkles in the deformed state. Therefore, we do not convert from the measure $\phi$ to the director orientation measured in the laboratory frame. This orientation $\phi$ is determined by the balance equations $\partial\mathcal{E}_\mathrm{s}/\partial\phi=0$, which yields
\begin{equation}
\tan2\phi=\frac{2 \varepsilon^{\mbox{\tiny(0)}}_{xy}}{\varepsilon^{\mbox{\tiny(0)}}_{xx}-\varepsilon^{\mbox{\tiny(0)}}_{yy}+h/\alpha}.\label{mbal}
\end{equation}
We suppose that due to the free boundary, $\sigma^{\mbox{\tiny(0)}}_{yy}=\sigma^{\mbox{\tiny(0)}}_{xy}=0$. Since the $\varepsilon^{\mbox{\tiny(0)}}_{xx}$ component of the strain is simply $-\gamma$, the imposed compression, equations (\ref{e2}--\ref{e3}) can be solved to give the remaining components of strain,
\begin{subequations}
	\begin{align}
\varepsilon^{\mbox{\tiny(0)}}_{xy}&=\frac{\alpha}{2 \mu} \sin(\phi) \cos (\phi),\label{es1}\\
\varepsilon^{\mbox{\tiny(0)}}_{yy}&= \frac{\gamma}{2}+\frac{\alpha}{4 \mu} \sin^2(\phi). \label{es2}
	\end{align}
\end{subequations}
 Substituting these expressions into the balance equation for the nematic director, Eq. \eqref{mbal}, allows us to calculate the director angle in terms of compressive strain and parameters of the NLCE. This equation admits the trivial solutions $\phi = 0$, $\phi = \pi/2$, corresponding respectively to the states $\mathbf{n}^{\mbox{\tiny(0)}} = \hat{\mathbf{x}}$ and $\mathbf{n}^{\mbox{\tiny(0)}} = \hat{\mathbf{y}}$, but also admits a third intermediate solution given by
\begin{equation}
\phi_\mathrm{c} = \frac{1}{2} \cos^{-1} \left[\frac{8h \mu -  \alpha ^2 - 12 \alpha \gamma \mu}{3 \alpha^2}\right]. \label{eq-critan}
\end{equation} 
By considering the regions of existence ($-1 \leq \mathrm{arg}(\phi_\mathrm{c}) \leq 1$)  for the three critical points $\phi=\left(0,\phi_\mathrm{c}, \pi/2\right)$, as well as their stability conditions,
\begin{equation}
 \partial ^2 \mathcal{E}_\mathrm{s}/\partial \phi^2 = -3 \alpha^2 \cos(4 \phi) - (\alpha^2-8 h \mu + 12 \alpha \gamma \mu) \cos(2 \phi) \geq 0, \label{stabsecond}
\end{equation}
we are able to construct a phase diagram of the base nematic state. The different phases (unrotated nematic director, rotating nematic director leading to an elastic soft mode, fully-rotated nematic director) are separated by two critical strains:
\begin{subequations}
	\begin{align}
\gamma_1&=\frac{2\mu h-\alpha^2}{3 \mu\alpha} ,\\
\gamma_2&= \frac{4\mu h+\alpha^2}{6 \mu\alpha},
	\end{align}
\end{subequations}
where for $\gamma<\gamma_1$ the director is unrotated, and for $\gamma>\gamma_2$ the director is fully-rotated. We show examples from this phase diagram in Fig.~\ref{tomographs}. Depending on the values of $(\alpha/\mu,h/\mu)$ at which the sheet was prepared, there are a few possible distinct behaviors of the director under small amounts of applied compression: no rotation, rotation only after a moderate critical strain, and an immediate jump in rotation under any amount of applied compression. In all cases, the director angle increases monotonically under the applied compression $\gamma$ and decreases monotonically in $h/\mu$. The behavior as a function of $\alpha/\mu$ is more complicated. For low values of $h/\mu$ (roughly $h/\mu<0.1$), there is a phenomenon of re-entrance as $\alpha/\mu$ varies, meaning that $\phi$ is not a monotonic function of $\alpha/\mu$, and holding $(\gamma,h/\mu)$ fixed whilst increasing $\alpha/\mu$ can take the director out of the rotating phase into the fully rotated phase, and then back into the rotating phase. The mechanism of this re-entrance is remarkably similar to that previously described for the nematic-smectic A re-entrant phase transition~\cite{pershanprost1979}, insofar as Eq. (\ref{stabsecond}) expanded in a Taylor series gives a fourth-order term in $\phi$ with coefficient $47 \alpha^2 + 8 h \mu - 12 \alpha \gamma \mu$, which for particular fixed values of $(h,\gamma,\mu)$ (the analog of temperature in the nematic-smectic A) can change signs as a function of $\alpha$ (the analog of pressure), which leads to re-entrance as the transition changes from second-order to first-order. 

\begin{figure}
\includegraphics[width=0.9\columnwidth]{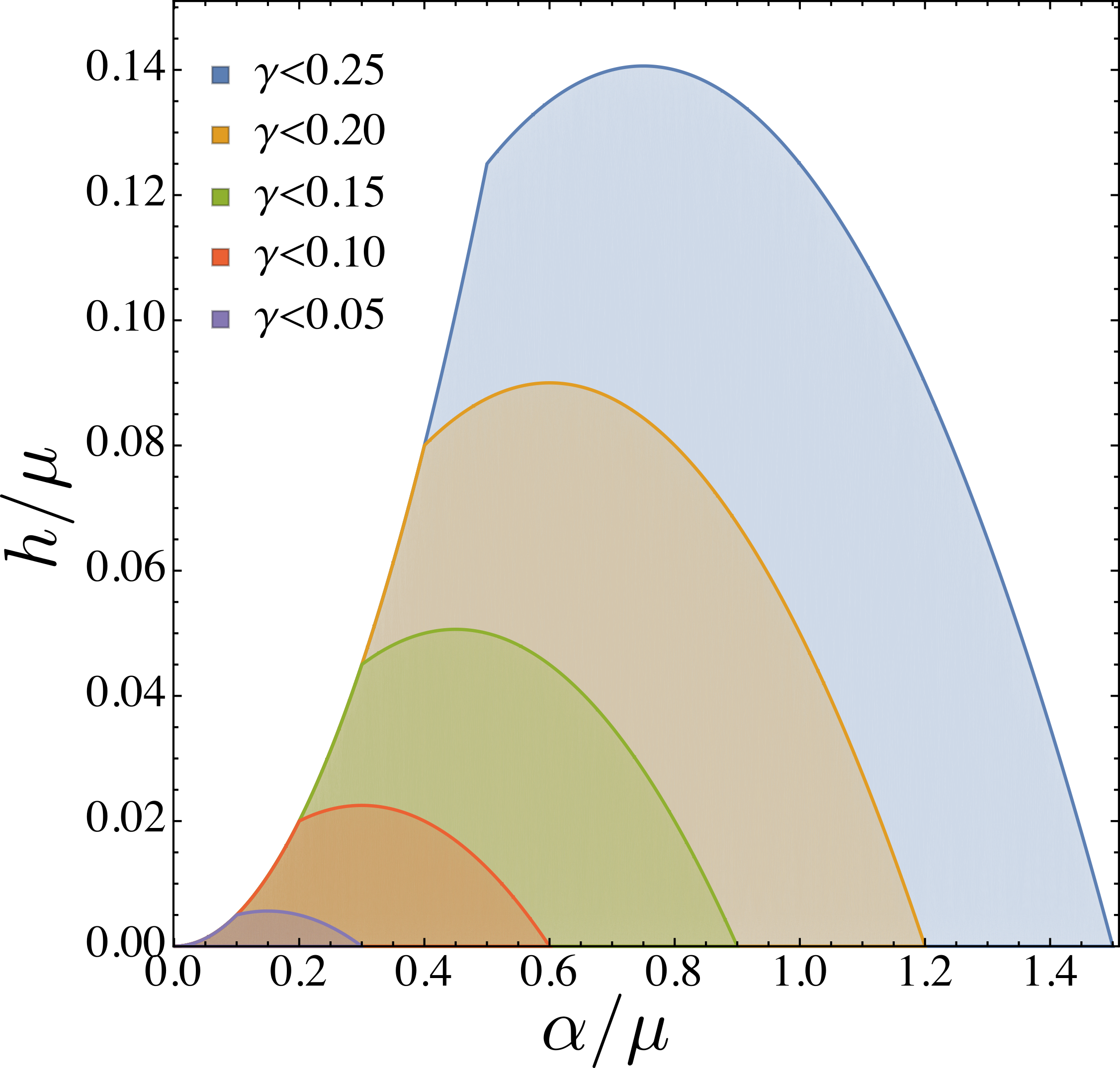}
\caption{\footnotesize Colored areas correspond to parameter values for which the stress in the film is extensile ($\sigma^{\mbox{\tiny(0)}}_{xx} >0$), which causes the sheet to remain flat no matter the compression of the substrate. This unusual situation arises from the presence of the soft mode of director rotation. Regions which are not colored either always have $\sigma^{\mbox{\tiny(0)}}_{xx} <0$, no matter the material parameters and compressive strain, or else require such a large compressive strain to complete the soft mode that they may be well beyond our small-strain theory.} 
\label{nowrinkles}
\end{figure}

When $\phi=\phi_c$, that is, the nematogen is neither unrotated nor fully rotated, we see the presence of a soft mode. This can be seen most easily by substituting Eq. \eqref{eq-critan} into Eq. \eqref{e1}, yielding 
\begin{equation}
\sigma^{\mbox{\tiny(0)}}_{xx}(\phi_c)=\alpha-2 \frac{h}{\alpha}\mu, \label{eq-softmodestress}
\end{equation}
which shows that the in-plane stress is effectively constant until the nematic finishes its rotation, even if the amount of compressive strain is increased. Additionally, this stress is not necessarily negative, an unusual feature in comparison to a classical elastomer. This gives a criterion for the suppression of wrinkling of a compressed plate which is completely independent of any substrate, because there can be no buckling if the stress is positive. In fact, such a plate would not be expected to buckle into a single arch in the absence of any substrate. We call this condition in which the plate would remain flat ``strong suppression'' of wrinkles: plates prepared with certain values of $(\alpha,h)$ would begin the soft mode (and also be in the regime where the soft mode leads to $\sigma^{\mbox{\tiny(0)}}_{xx}>0$) under any amount of compressive strain, and would not complete the soft mode (i.e., rotate a total of $\pi/2$ from the base state) until a certain amount of strain was imposed, after which time buckling may be possible. Some examples are plotted in Fig.~\ref{nowrinkles}. 

\section{Out-of-plane displacement, critical buckling strain, and scaling of wrinkles}
\label{outplanesec}
We now assume an out-of-plane displacement which depends only on the in-plane coordinate $x$, $\zeta(x)=A \cos (q x)$ (we perform an analogous calculation for a plate of finite width $W$ and out-of-plane displacement $\zeta(x,y)$  in the Appendix, in which we demonstrate that identical results hold in comparison to this infinite, one-dimensional problem). Our particular choice of geometry allows us to simplify the bending energy \eqref{bendingenergy}. The bending energy per unit wavelength (and per unit width, so effectively an energy per unit area) reads
\begin{equation}
\mathcal{E}_\mathrm{b} = \frac{t^3}{24} \frac{q}{2 \pi}  \int _0 ^{2 \pi/q} \mathrm{d}x  \,\mathcal{B}(\phi) \left(\partial^2_{x}\zeta\right)^2 = \frac{t^3 \mathcal{B}(\phi) q^4 A^2}{48}, \label{bendingenergysimp}
\end{equation}
where the renormalized Young's modulus is given by
\begin{equation}
\mathcal{B}(\phi) = \!4 \mu\!-\!\frac{\alpha}{2} \left[3\!+\!5 \cos (2 \phi)\right]\!+\!\frac{h}{2}\left[5\!+\!3 \cos (2 \phi )\right] \cos^2\phi.
\end{equation}
Following a standard calculation~\cite{CerdaMahadevan2003,Davidovitch2011,HuangHongSuo05, AudolyBoudaoud2008}, we write the stretching energy density as 
\begin{eqnarray}
\mathcal{E}_\mathrm{s} &=&  \frac{t}{2}  \  \sigma^{\mbox{\tiny(0)}}_{xx} \left(\partial_x\zeta\right)^2, 
\end{eqnarray}
where $\sigma^{\mbox{\tiny(0)}}_{xx}$ is given by Eq. \eqref{e1}, and can be written in the form $\sigma^{\mbox{\tiny(0)}}_{xx} = -3\mu \gamma + (3/2)\alpha\sin ^2 \phi$. While there are other contributions to the stretching energy, they are independent of the wavelength of wrinkling and will therefore not contribute to the variational problem which gives this wavelength. For instance, following \cite{HuangHongSuo05}, the total strain perpendicular to the wrinkles can be written as $\varepsilon^{\mbox{\tiny(0)}}_{xx}=-\gamma+\frac{1}{4} q^2 A^2$, so that by Eq. \eqref{e1} the addition to the stress is simply $\mu q^2 A^2$. The stretching energy per unit area (keeping only terms proportional to $q$) is therefore
\begin{eqnarray}
\mathcal{E}_\mathrm{s} &=&  \frac{t}{4} (\mu q^4 A^4 + \sigma^{\mbox{\tiny(0)}}_{xx} q^2 A^2 ). \label{exstretchen}
\end{eqnarray}
Finally, we assume that the thin NLCE film is bonded to an underlying foundation~\cite{landau_lifshitz_elas,CerdaMahadevan2003}, which can be either fluid or elastic. Because the energy of the foundation depends slightly differently on the wrinkling in each case---the elastic foundation energy depends both on the amplitude and the wavelength of wrinkles, whereas the fluid foundation energy depends only on the amplitude---we will treat each case separately below.

\begin{figure*}
\includegraphics[width=1.9\columnwidth]{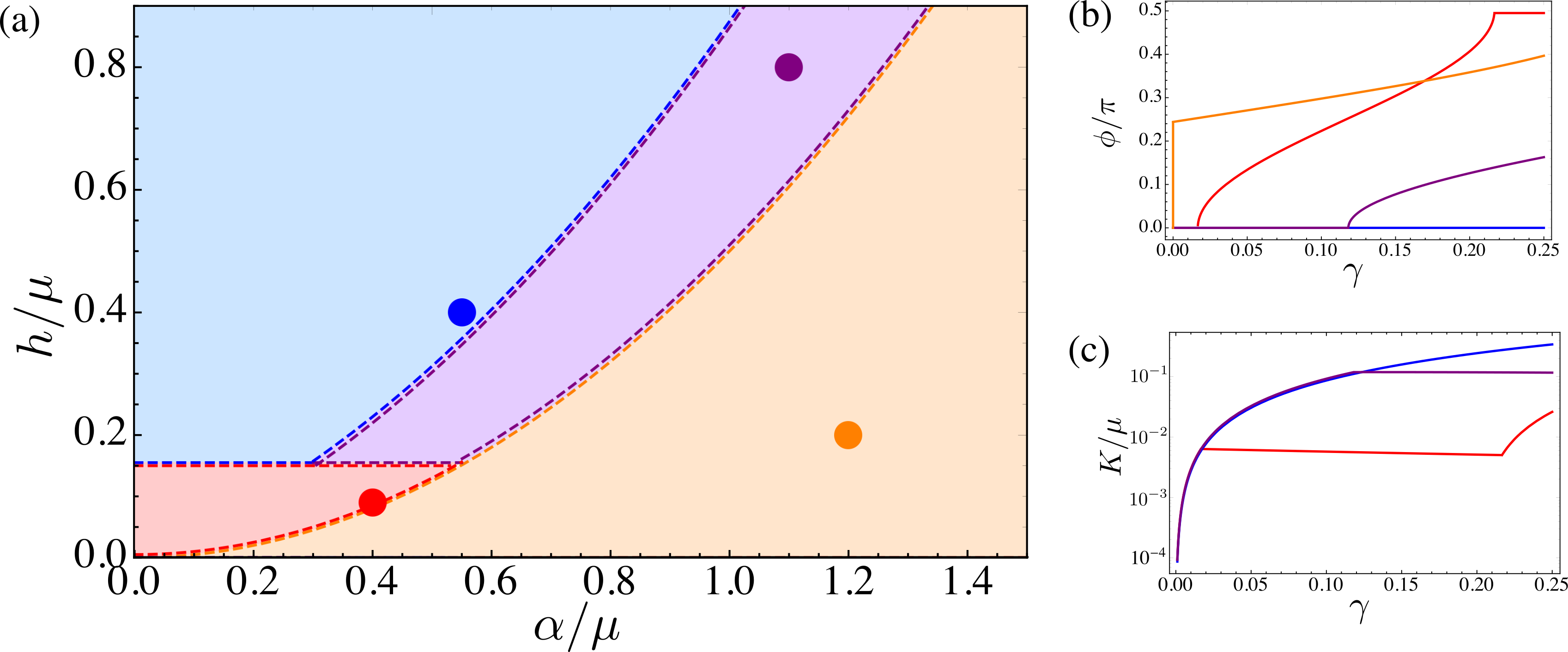}
\caption{\footnotesize Variation of the critical strain as a function of strain-director coupling and crosslinking strength. (a) Position of example values of $(\alpha,h)$ used, and regions where the example used represents the qualitative behavior of other plates. (b) The internal degree of freedom (nematic director angle) $\phi$ as a function of the applied compression $\gamma$, which does not depend on an underlying substrate. (c) The dependence of the critical strain for buckling (here, shown on the $x$-axis for convenience) on the Young's modulus $K$ of the underlying substrate, depends drastically on the internal degree of freedom as shown in (b). For instance, the blue plate never rotates, and obeys the classical buckling law; the orange plate remains entirely in the soft mode (and in the region of positive stress for $\gamma$ less than a very large value, even larger than those plotted in Fig.~\ref{nowrinkles}; therefore, it cannot buckle, and is not plotted here) for the whole range of $\gamma$ shown; the purple plate remains classical until it begins to rotate at $\gamma \approx 0.11$; the red plate goes through all three phases (unrotated, soft, fully-rotated). During the soft mode, the critical strain for buckling (when applicable) is highly insensitive to $K$.}
\label{fig-critstrain}
\end{figure*}

\subsection*{Elastic substrate}

Following previous derivations of the energetics for an elastic sheet atop an infinitely deep substrate of Young's modulus $K$~\cite{ChenHutchinson2004,HuangHongSuo05}, the wrinkling induces a normal stress $\bar{\sigma}_{zz}=K q \zeta$, so that the energy per area stored in the foundation can be written as
\begin{equation}
\mathcal{E}_\mathrm{e}= \frac{q}{2 \pi} \int _0 ^{2\pi/q} \mathrm{d}x \, \frac{1}{2} K q \zeta^2 = \frac{1}{4} K q A^2, \label{elfund}
\end{equation}
and the total energy in the system (again, neglecting terms not proportional to $q$) is, collecting in terms of powers of $q$,
\begin{equation}
4 \mathcal{E}= t \mu q^4 A^4 + q^2 A^2 \left( \frac{t^3 \mathcal{B}(\phi) q^2}{12} + t \sigma^{\mbox{\tiny(0)}}_{xx} + \frac{K}{q} \right). \label{eltot}
\end{equation}
When viewed as a function of the wrinkling amplitude $A$, it is clear that wrinkling can only occur when the term inside the parentheses is negative, or else the optimal amplitude will always be zero. It is convenient to write this term as $ t \sigma^{\mbox{\tiny(0)}}_{xx} + \mathcal{B}(\phi) t \xi_\mathrm{e}$, where
\begin{equation}
\xi_\mathrm{e}=\frac{(t q)^2}{12}+\frac{K}{\mathcal{B}(\phi) q t}. \label{xie}
\end{equation}
When the thin plate is wrinkled, the energy is then minimized at an amplitude of 
\begin{equation}
|A|= \frac{1}{q} \sqrt{\frac{-( \sigma^{\mbox{\tiny(0)}}_{xx}+\mathcal{B}(\phi) \xi_\mathrm{e})}{2 \mu}} \label{elamp}. 
\end{equation}
Therefore the plate can only wrinkle if $|\sigma^{\mbox{\tiny(0)}}_{xx}|> \mathcal{B}(\phi) \xi_\mathrm{e}$, and the stress must also be negative (compressive), which is not always the case, as seen in Sec. \ref{inplanesec}. Substituting Eq. \eqref{elamp} into Eq. \eqref{eltot} and minimizing the energy gives the wavelength $\lambda_\mathrm{e}$ of wrinkling on an elastic substrate,
\begin{equation}
\lambda_\mathrm{e} =\frac{ 2 \pi}{q_\mathrm{e}}= 2 \pi \,\ell_\mathrm{e} \left(\frac{\mathcal{B}(\phi)}{\mu}\right)^{1/3}, \label{elastscale}
\end{equation}
where $\ell_\mathrm{e}\equiv t[\mu/(6K)]^{1/3}$ is the natural length emerging in this problem. Eq. \eqref{elastscale} allows us to write the critical stress for wrinkling in the case where the stress remains compressive,
\begin{equation}
|\sigma^{\mbox{\tiny(0)}}_{xx}|= \frac{1}{2} \left(3 K  \right)^{2/3} \mathcal{B}^{1/3} . \label{elastcrit}
\end{equation}
On their face, Eqs. \eqref{elastscale} and \eqref{elastcrit} are identical to those found in the literature for a thin, non-nematic elastomer plate of infinite extent across an elastic foundation of infinite depth~\cite{ChenHutchinson2004,HuangHongSuo05,AudolyBoudaoud2008}. However, the renormalized Young's modulus $\mathcal{B}(\phi)$ as well as the in-plane stress $\sigma^{\mbox{\tiny(0)}}_{xx}$ now carry information from the embedded nematic phase --- therefore, tuning parameters relevant to the nematic can both change the wavelength of wrinkles at a fixed amount of compression, $\gamma$, and can also change the threshold for buckling. 

To demonstrate the interesting ways in which information from the nematic phase can affect the critical buckling strain of the entire sheet, we have plotted information about four representative elastomer plates in Fig.~\ref{fig-critstrain}. These four plates were chosen because they represent the four essential behaviors of the director angle which are seen in Fig.~\ref{tomographs} if the applied compressive strain never exceeds $\gamma=0.25$: no rotation (high $h$, low $\alpha$); a large soft mode characterized by rotation across the whole range of $\gamma$ (low $h$, high $\alpha$); rotation which begins only after a moderate applied strain (high $h$, high $\alpha$); and a plate which both begins and ends its rotation within the range of $\gamma$ used (low $\alpha$, low $h$). The curved lines in Fig.~\ref{fig-critstrain}(a) delimit the regions where the plate will never rotate for $\gamma<0.25$ (blue-purple) and where the soft mode induces positive stress (purple-orange). The flat line (blue-red) roughly delimits plates which begin and end rotation before $\gamma=0.25$. During the rotation of the director, the critical buckling strain becomes almost completely independent of the Young's modulus of the underlying substrate, which acts to suppress wrinkling. For instance, the red plate bonded to a substrate with $K/\mu=0.01$ has a critical buckling strain of roughly $\gamma=0.22$, whereas a blue plate atop the same substrate has a critical buckling strain of roughly $\gamma=0.03$.

\begin{figure}[tbh]
\centering
\includegraphics[width=0.95\columnwidth]{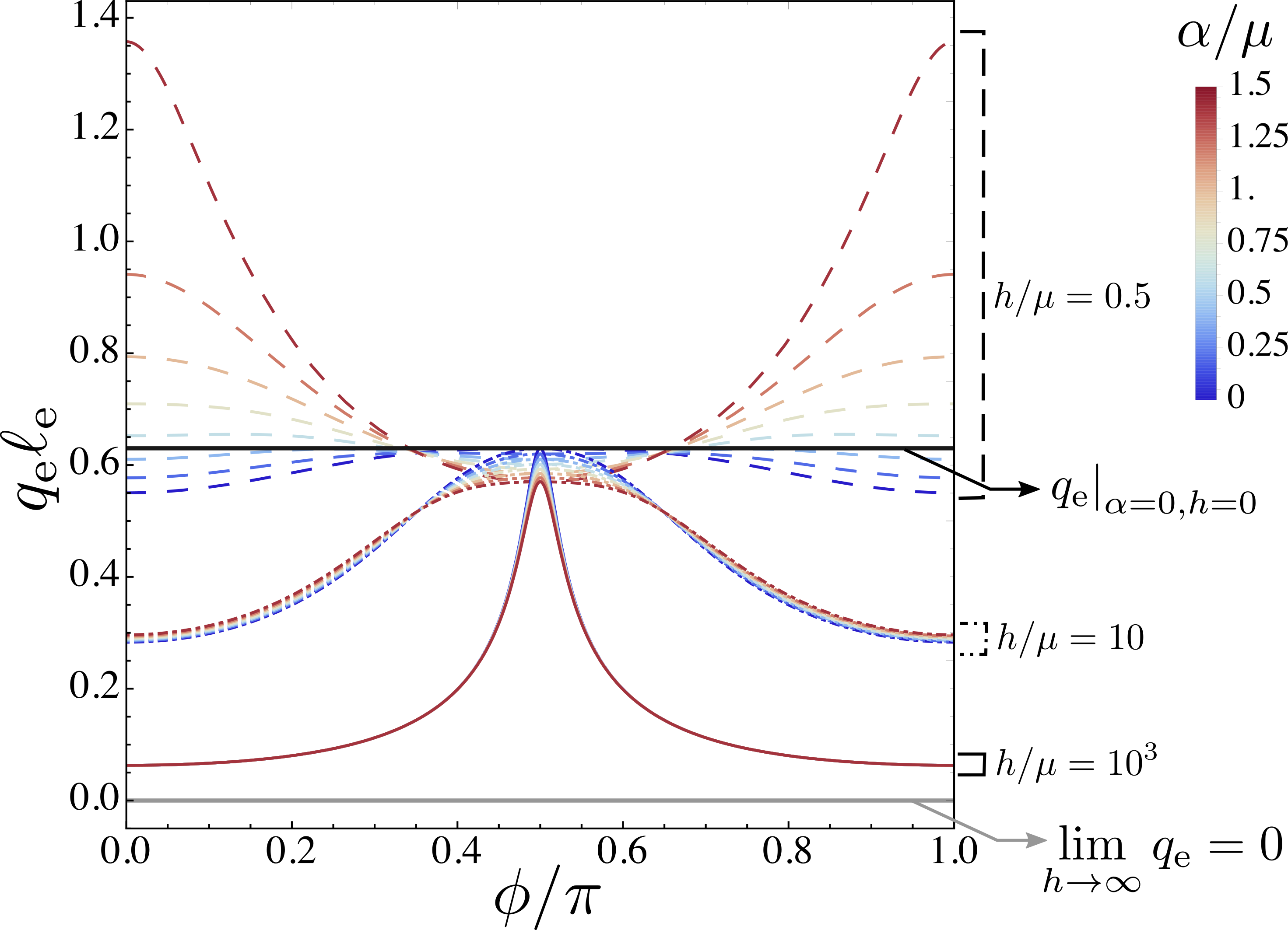}
\caption{\footnotesize Wave number of wrinkles on an elastic foundation, normalized by the natural length $\ell_\mathrm{e}\equiv t[\mu/(6K)]^{1/3}$, versus $\phi/\pi$. Different groups for the same value of $h/\mu$ (dashed $h/\mu=0.5$, dotted-dashed $h/\mu=10$, and solid $h/\mu=10^3$) refer to the strength of the anchoring term, whereas the color map represents the strength of the elasticity-order coupling, $0\leq \alpha/\mu \leq 3/2$. The solid black line gives the wavelength when there is no nematic, $\alpha/\mu=0$ and $h/\mu=0$, and the gray line sets the bound for a nematic glass, \emph{i.e.} $h \rightarrow \infty$.} 
\label{weakdep}
\end{figure}

\subsection*{Fluid substrate}

Atop a fluid substrate, the increase in the energy density is due to hydrostatic pressure, which has an energetic cost given by
\begin{equation}
\mathcal{E}_\mathrm{f}= \frac{1}{2} \int \mathrm{d}x \, \rho g \zeta^2 = \frac{1}{4} \rho g A^2, \label{flfund}
\end{equation}
where $\rho$ is the density of the fluid and $g$ is the gravitational field. The analogue of the quadratic term in Eq. \eqref{xie} is therefore 
\begin{equation}
\xi_\mathrm{f}=\frac{(q t)^2}{12} + \frac{\rho g}{t \mathcal{B}(\phi) q^2}. \label{xif}
\end{equation}
Eq. \eqref{elamp} and the critical stress for wrinkling are therefore unchanged (save for the substitution of $\xi_\mathrm{f}$ for $\xi_\mathrm{e}$); however, because of the extra factor of $q^{-1}$ in $\xi_\mathrm{f}$, the critical wavelength is slightly different: 
\begin{eqnarray}
\lambda_\mathrm{f} &=& \frac{2\pi}{q_\mathrm{f}}=2 \pi \,\ell_\mathrm{f}\left(\frac{\mathcal{B}(\phi)}{\mu} \right)^{1/4}, \label{fluidscale}
\end{eqnarray}
where $\ell_\mathrm{f}\equiv[t^3\mu/(12 \rho g)]^{1/4}$ is the natural length associated with the cost of deforming the fluid substrate. This expression is again identical to the equivalent expression for a regular, isotropic elastomer, aside from the fact that $\mathcal{B}(\phi)$ carries information from the nematic orientation, which is dictated by the in-plane problem. Therefore, the wavelength of the wrinkles or even their existence is highly dependent upon the base state of stress, much unlike the  case of an isotropic elastomer. The critical stress, again in the case where the stress remains compressive, is
\begin{equation}
|\sigma^{\mbox{\tiny(0)}}_{xx}|= \sqrt{\frac{\rho g t  \mathcal{B}(\phi)}{3 }}, \label{fluidcrit}
\end{equation}
much of which can still be inferred from Fig.~\ref{nowrinkles}, though now the critical stress is dependent on the thickness $t$ of the elastomer plate.

We note that in addition to the region described in Fig.~\ref{nowrinkles} where wrinkling is forbidden due to a change from compressive to extensile stress, wrinkling is also suppressed for large $h$ due to effective stiffening of the sheet. We show this in Fig.~\ref{weakdep} for the case of an elastic substrate, although a nearly-identical result holds for the fluid substrate as well. Because this limit corresponds to a nematic glass~\cite{CirakWarner2014,Lubensky_etal2002}, the mechanism for instability suppression in this region is the strong coupling between curvature and defects in the nematic texture~\cite{modes_warner2011}. In Eq.~(\ref{bendingenergy}), we see that for $h \gg \alpha$ the bending energy is very large unless $B_{\alpha\beta}^2$ and $C_{\alpha\beta}\Delta{Q}^{\mbox{\tiny(0)}}_{\alpha\beta}$ are very small. Since our textures are free of defects, the presence of any curvature is highly penalized for arbitrarily-increasing $h$. 

\section{Summary}
\label{sumsec}
In this work we derived a F\"{o}ppl-von K\'{a}rm\'{a}n type of plate theory for a thin NLCE which was cross-linked deep in the nematic phase. This was accomplished following the standard techniques of dimensional reduction for thin elastic bodies, employing the additional assumption that the nematic director was tangent to the mid-plane before and after the deformation. In the limit of small and large $h$, the coupling parameter for the ``memory'' or ``anchoring'' term corresponding to the fact that the elastomer was cross-linked in the nematic phase, our model corresponds to other objects studied in the literature: respectively, an elastomer cross-linked in the isotropic phase ($h \rightarrow 0$)~\cite{Lubensky_etal2002,WarnerTerentjev} or a nematic glass ($h \rightarrow \infty$)~\cite{bigginswarner2008,CirakWarner2014,wmc2010,modes_warner2011}. When both this coupling and the director-strain coupling parameter vanish, we recover the equations for the isotropic plate.

The model is fairly simple and solutions to certain geometries and boundary conditions can be found analytically. As a first step in this direction, we calculated the wavelength of wrinkles for a compressed NLCE plate atop a fluid or elastic foundation. The wave number is found to be non-monotonic in the compressive strain $\gamma$ and highly dependent on the nematic-elastomer coupling parameters, as it is shown in Fig.~\ref{weakdep}. For certain sets of coupling parameters the plate does not wrinkle until a very high strain threshold is reached. We therefore believe it will be possible to design NLCE which are not subject to the elastic instabilities of classical rubbers until a prescribed strain threshold is met. 

In summary, NLCE  plates present a new venue to study nontrivial pattern formation in thin sheet elasticity and also present an example of robust control of mechanical stability through the coupling of geometry and microstructure. We hope that our simplified model will be useful in further study of liquid crystal elastomers.

\begin{acknowledgments}
We are grateful to Thomas Powers for his guidance and the many discussions had concerning this project. We would also like to thank Badel Mbanga, Timothy Atherton, James Hanna, and Jonathan Selinger for helpful conversations.
\end{acknowledgments}

\appendix*
\section{Wrinkling of a plate with finite width}
\label{appsec}

\setcounter{figure}{0}
\makeatletter 
\renewcommand{\thefigure}{A\@arabic\c@figure}
\makeatother

\begin{figure}
\includegraphics[width=0.8\columnwidth]{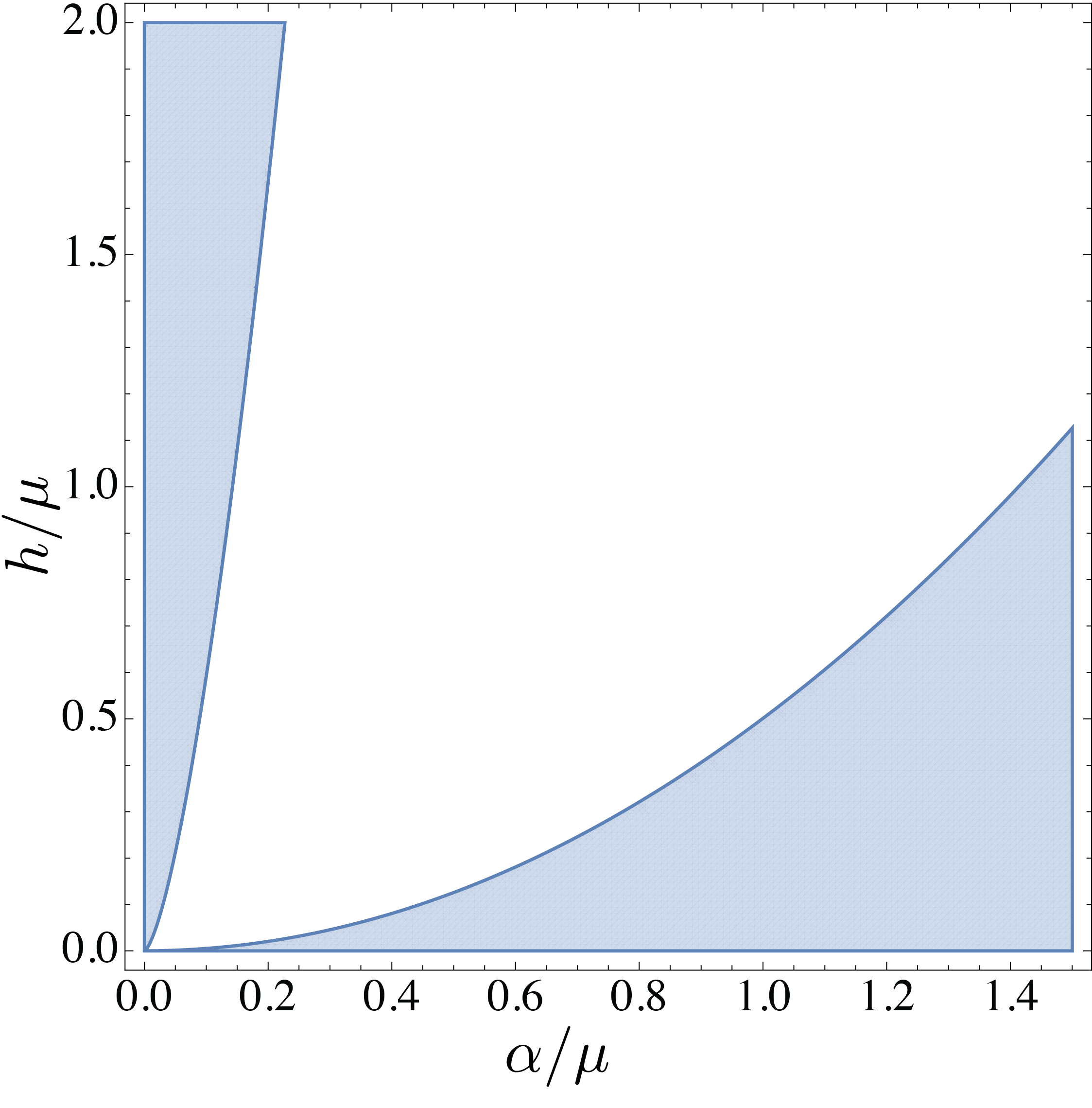}
\caption{\footnotesize Areas of parameter space in the $\gamma=0.1$ plane for which $\omega ^2 < D^2$ (blue) and therefore $\zeta(x,y)=0$. Here as an example we take $t = 1\,\mathrm{mm}$, $\rho g/\mu = 0.05$, representative of a rubber film atop a water substrate. The full finite two-dimensional case has a plane-stress condition which is weakly dependent on the wavenumber of wrinkling; we choose $q=8 mm$ to mimic the wavenumber in an elastomer that is not nematic. Because here $\omega^2 < D^2$ represents one condition rather than the two described in the main text (the stress must be (a) compressive and (b) beyond a certain threshold), we plot both regions (so that the region at left, which does not appear in Fig.~\ref{nowrinkles}, corresponds to a region where the stress is still compressive, but insufficient to cause wrinkling in the plate with the parameters given).} 
\label{nowrinklessupp}
\end{figure}

When we consider a plate of finite width, the stress is the same as found in \eqref{e1}-\eqref{e3}, but we allow for out-of-plane deformations which depend on both spatial coordinates, $\zeta(x,y)$. For the sake of convenience, we consider here the case of a sheet which has one small lateral dimension: $0 < y < W$, where $ W \ll L$. Following previous work on wrinkling in a similar geometry~\cite{CerdaMahadevan2003} we reason that because the out-of-plane displacement varies much less in $y$ than in $x$ the respective terms in the energy are also smaller: $\left(\partial_y^2\zeta\right)^2 < \left(\partial_x\partial_y\zeta\right) \left(\partial_y^2\zeta\right) \ll \left(\partial_{x}^2\zeta\right)^2$; we will consequentially choose to keep only the largest terms. The total bending energy reads
\begin{eqnarray}
&&E_\mathrm{b}= \frac{ t^3}{24}  \int_A \left[\mathcal{B}^{xxxx} \left(\partial_{x}^2\zeta\right)^2 + \mathcal{B}^{xyxy} \left(\partial_x\partial_y\zeta\right)^2  +\right. \nonumber\\ 
&&\!\!\!\!\!+\left.\mathcal{B}^{xyxx} \! \left(\partial_x\partial_y\zeta\right)\!  \left(\partial_x^2\zeta\right)\!\! +\!\! \mathcal{B}^{xxyy} \!\left(\partial_x^2\zeta\right) \!\left(\partial_y^2\zeta\right)\right]\mathrm{d}\mathcal{A}^{\mbox{\tiny(0)}},
\end{eqnarray}  
where the respective bending stiffnesses are given by
\begin{subequations}
	\begin{align}
\mathcal{B}^{xxxx} &= 4 \mu-\frac{\alpha}{2} \left[3+5 \cos (2 \phi)\right]+\frac{h}{2}\left[5+3 \cos (2 \phi )\right] \cos^2\phi\\
\mathcal{B}^{xyxy} &= 4 \mu-4 \alpha +\frac{h}{2}\left[6+\cos (2 \phi )-3 \cos(4 \phi )\right] \\
\mathcal{B}^{xyxx} &= -5 \alpha \sin (2 \phi )+h\left[4+3 \cos (2 \phi )\right]\sin (2 \phi )\\
\mathcal{B}^{xxyy} &= 4 \mu+\frac{3}{2} h \sin ^2(2 \phi ).
	\end{align}
\end{subequations}
Note that $\mathcal{B}^{xxxx}$ corresponds to $\mathcal{B}(\phi)$ in \eqref{bendingenergysimp}. Because the stress and foundation energy remain unchanged, we have now the balance equation
\begin{eqnarray}
0 &=&  \frac{t^3}{12} \left(\mathcal{B}^{xxxx} \partial _x ^4 \zeta + (\mathcal{B}^{xyxy} + \mathcal{B}^{xxyy}) \partial _x ^2 \partial _y ^2 \zeta + \mathcal{B}^{xyxx} \partial_x ^3 \partial _y \zeta\right) \nonumber \\
&&+ 3 t \left( \mu \gamma - \frac{1}{2} \alpha\sin ^2 \phi\right) \partial _x ^2 \zeta  + \frac{\delta E_{\mbox{\tiny(f,e)}}}{\delta \zeta}.
\label{mahabal3}
\end{eqnarray}
We assume that the out-of-plane displacement takes the form $\zeta =   e^{(i q x)} Y (y)$. That is, the wrinkles are cylindrical as before, but the function $Y (y)$ will serve to implement the free boundary conditions at the edges. This gives a  Sturm-Liouville equation for $Y(y)$:
\begin{equation}
0 = A_1 Y_n + i A_2 Y'_n + A_3 Y''_n,
\end{equation}
where
\begin{eqnarray}
A_1 &=&  \frac{q^4t^3\mathcal{B}^{xxxx}}{12}+\frac{\partial^2 E_{\mbox{\tiny(f,e)}}}{\partial \zeta^2}-q^2t (3 \mu \gamma + \frac{3 \alpha}{2}(\cos ^2 \phi - 1)), \nonumber \\
A_2 &=& - \frac{q^3t^3}{12} {h}  , \nonumber \\
A_3 &=&  -\frac{q^2 t^3}{12}(\mathcal{B}^{xyxy}+\mathcal{B}^{xxyy}).
\end{eqnarray}
Upon substitution of the \emph{ansatz} $Y(y) = A e^{\kappa y}$, the characteristic equation is obtained:
\begin{equation}
\kappa ^2 +  2 D \kappa + \omega ^2 = 0,
\end{equation}
where $\omega  ^2 = A_1/A_3$ and $2 D  = iA_2/A_3$. This equation has the roots $\kappa  = -D  \pm \sqrt{D  ^2 - \omega  ^2}$. As a consequence, $Y(y)$ is either an exponential function or a sinusoid; the former must be disbarred by symmetry, so that $Y(y) = 0$ whenever $\kappa \in \mathbb{R}$. Because $A_2, \ A_3$ are real, $D$ satisfies $D ^2 < 0$. Therefore, wrinkling solutions are only valid when $\omega^2 \geq D^2$. 

The scaling for the finite two-dimensional sheet is the same as in the one-dimensional cylindrical approximation. To show this, we consider the sinusoidal solution for $Y(y)$:
\begin{equation}
Y_n =  A \sin[Im(\kappa ^+)y] + B \sin[Im(\kappa ^-)y]
\end{equation}
where $\kappa^+, \ \kappa^-$ indicates which sign of the $\pm$ in the root is being considered. Minimizing the bending energy allows us to simplify this expression to $Y_n (y) = \sin\left(\pi y/W\right)$, because additional modes increase the bending energy prohibitively. Integrating the total energy and keeping only terms algebraic in $q$ yields
\begin{gather}
\frac{E_\mathrm{T}}{LW \varepsilon^{\mbox{\tiny(0)}}_{yy}}= \left(\frac{1}{q^2} \frac{\delta^2 E_{\mbox{\tiny(f,e)}}}{\delta \zeta^2}  + \frac{1}{4} t \sigma^{\mbox{\tiny(0)}}_{xx}\right) \nonumber \\
+\frac{t^3}{12}\left( \mathcal{B}^{xxxx} q ^2 + \frac{\pi^2}{W^2}(\mathcal{B}^{xxyy}+\mathcal{B}^{xyxy}) \right). 
\end{gather}
Because the new contributions to the bending energy which are proportional to other bending moduli do not scale with $q$ and the foundation and stretching energies remain unaltered, the scalings given in \eqref{fluidscale}, \eqref{elastscale} are still applicable for the fully finite 2D sheet with saddle-shaped wrinkles. 

\bibliography{references.bib}

\end{document}